\documentclass[12pt]{article}
\usepackage{amsfonts}

\usepackage{amsmath}

\csname @addtoreset\endcsname{equation}{section}
\newcommand{\eq}{\begin{equation}}
\newcommand{\feq}{\end{equation}}
\newcommand{\eqn}{\begin{eqnarray}}
\newcommand{\feqn}{\end{eqnarray}}
\newcommand{\arr}{\begin{eqnarray*}}
\newcommand{\farr}{\end{eqnarray*}}

\font\mybb=msbm10 at 12pt
\def\bb#1{\hbox{\mybb#1}}
\def\bZ {\bb{Z}}
\def\bR {\bb{R}}

\def\bC {\bb{C}}

\begin{document}

\begin{titlepage}
\begin{flushright}
IFUM 657/FT\\
CAMS/00-04\\
hep-th/0004077
\end{flushright}
\vspace{.3cm}
\begin{center}
\renewcommand{\thefootnote}{\fnsymbol{footnote}}
{\Large \bf Entropy of Black Holes in $D=5$, $N=2$ Supergravity and
AdS Central Charges}
\vfill
{\large \bf {S.~Cacciatori$^1$\footnote{email: cacciatori@mi.infn.it},
D.~Klemm$^1$\footnote{email: dietmar.klemm@mi.infn.it},
W.~A.~Sabra$^2$\footnote{email: ws00@aub.edu.lb},
and D.~Zanon$^1$\footnote{email: daniela.zanon@mi.infn.it}}}\\
\renewcommand{\thefootnote}{\arabic{footnote}}
\setcounter{footnote}{0}
\vfill
{\small
$^1$ Dipartimento di Fisica dell'Universit\`a di Milano
and\\ INFN, Sezione di Milano,
Via Celoria 16,
20133 Milano, Italy.\\
\vspace*{0.4cm}
$^2$ Center for Advanced Mathematical Sciences (CAMS)
and\\
Physics Department, American University of Beirut, Lebanon.\\}
\end{center}
\vfill
\begin{center}
{\bf Abstract}
\end{center}
{\small We consider general black holes in $D=5$, $N=2$
supergravity coupled to vector multiplets, and discuss the issue
of microstate counting from various viewpoints.
The statistical entropy is computed for the near-extremal case using the
central charge of the $AdS_2$ factor appearing in the near-horizon geometry.
Furthermore, we explicitly construct the
duality transformation connecting  electrically charged black holes to 
magnetically
charged black strings, under which the $AdS_2\times S^3$ near horizon 
geometry
becomes $AdS_3\times S^2$.
For $AdS_3$ the counting of microstates correctly reproduces the
Bekenstein-Hawking entropy, thus resolving the discrepancy previously
found for $AdS_2$.
}

\end{titlepage}

\section{Introduction}

\label{intro}

The study of black hole solutions in $N=2$ five-dimensional supergravity
coupled to vector and hypermultiplets plays an important role in the
understanding of the non-perturbative structure of string and M-theory \cite
{chou,gaida}. In this setting the interplay between classical and quantum
results is exemplified at its best.

In this paper we consider general charged black holes of the $D=5$, $N=2$
theories, not necessarily those obtained from compactification
of eleven-dimensional supergravity on a Calabi-Yau threefold. The analysis
is simplified by the rich geometric structure of the $N=2$ theories. Black
hole solutions are given in terms of a rescaled cubic homogeneous
prepotential which defines very special geometry \cite{antoine}. In the
extremal BPS case, half of the vacuum supersymmetries are preserved, while
at the horizon supersymmetry is fully restored \cite{chamsabra2}.

In five dimensions the supergravity action contains a Chern-Simon term which
allows the existence of black holes with nonvanishing angular momenta (but
still nonrotating horizon) \cite{cvetic1,bmpv,krw,chamsabra1}.

These issues have been the object of recent, extensive studies. In the present
paper we focus on the asymptotic symmetries of the near-horizon geometry of
the general near-extremal solution: the aim is the computation of the
entropy from a counting of microstates to be compared to the macroscopic,
thermodynamical entropy.

We will see that the calculation of the microscopic entropy of small
excitations above extremality is equivalent to a microstate counting for
certain black holes in two-dimensional anti-de~Sitter space. The latter,
however, is problematic: $AdS_2$ has two timelike boundaries, but when
applying Cardy's formula for the density of states only one boundary is
taken into account. This procedure leads to a statistical entropy result
which is off by a $\sqrt{2}$ factor with respect to the Bekenstein-Hawking
entropy \cite{cadmig}. Up to now, no satisfactory explanation of this
mismatch is known. Our results support the
point of view that only the ground state has an effective description
in terms of a quantum-mechanical system \cite{fragm}, whereas
the excitations above extremality are described by a two-dimensional
conformal field theory \cite{maldstrom}.

In our work we address these issues in a constructive approach. The main
result that we present is an explicit duality transformation, which realizes
an invariance of the $N=2$ supergravity action. This duality turns the $%
AdS_{2}\times S^{3}$ near horizon geometry of the extremal black hole
solution into $AdS_{3}\times S^{2}$. The key point underlying the duality is
the fact that the three-sphere can be written as a Hopf fibration over the
base $S^{2}$. For $AdS_{3},$ the counting of microstates is performed using
Cardy's formula and it is shown that it reproduces correctly the
Bekenstein-Hawking entropy, thus resolving the discrepancy previously found
for $AdS_{2}$.

In the case where the $D=5$, $N=2$ supergravity action is obtained by
Calabi-Yau (CY) compactification of M-theory, the considered duality
transformation, which maps electrically charged black holes onto
magnetically charged black strings, corresponds to the duality between M2
branes wrapping CY two-cycles and M5 branes wrapping CY four-cycles.
According to \cite{maldacena}, M-theory compactified on $AdS_{3}\times
S^{2}\times M$, where $M$ denotes some Calabi-Yau threefold, is dual to a $%
(0,4)$ superconformal field theory living on an M5 brane wrapping some
holomorphic CY four-cycle. This fact has been used in \cite{vafa} to compute
the entropy of five-dimensional BPS black holes\footnote{%
The work in \cite{vafa} includes as a special case also the results obtained
in \cite{stromvafa}.}. We stress that our method for microstate counting
applies to any near-extremal black hole in $N=2$, $D=5$ supergravity,
independent of whether it is obtained by CY compactification or not.

Our paper is organized as follows: in section 2, the basic notions of $N=2$,
$D=5$ supergravity and very special geometry relevant to our analysis are
summarized. In section 3 we review the black hole solutions and consider the
$STU$ model as a simple example, which nonetheless
retains all the interesting features of the general solutions. In section 4
we present the isometry superalgebra which arises in the near horizon limit,
while in section 5 we show that the motion of a particle which moves near
the horizon of the extremal rotating black hole is described by  conformal
quantum mechanics. This indicates that the ground state may have a
description in terms of conformal quantum mechanics \cite{claus,gibbtown},
even when rotation is included. In section 6 we compute the statistical
entropy of small excitations near extremality, using the $AdS_{2}$ central
charge \cite{cadmig}, and find a $\sqrt{2}$ factor of discrepancy as
compared to the thermodynamical Bekenstein-Hawking result. In section 7 we
construct the duality transformation for the supergravity action, and in the
following section we finally perform the state counting, using the fact that
the near-horizon geometry of the dual solution includes an $AdS_{3}$ factor.
In this way, we obtain a microscopic entropy which agrees precisely with the
corresponding thermodynamical result. We conclude with some final remarks.

\section{$D=5$, $N=2$ Supergravity and Very Special Geometry}

The theory of $N=2$ supergravity theory coupled to an arbitrary number $n$
of Maxwell supermultiplets was first considered in \cite{GST}. In the
analysis of \cite{GST}, it was established that the scalar fields of the
vector multiplets parametrize a Riemannian space. The homogeneous symmetric
spaces take the form

\begin{equation*}
\mathcal{M}=\frac{Str_{0}(J)}{Aut(J)},
\end{equation*}
where $Str_{0}$($J)$ is the reduced structure group of a formally real
unital Jordan Algebra of degree three, $Aut(J)$ is its automorphism group.

The scalar manifold can be regarded as a hypersurface, with vanishing second
fundamental form of an $(n+1)$-dimensional Riemannian space $\mathcal{G}$
whose coordinates $X$ are in correspondence with the vector multiplets
including that of the graviphoton. The equation of the hypersurface is $%
\mathcal{V}=1$ where $\mathcal{V}$, the prepotential, is a homogeneous
cubic polynomial in the coordinates of $\mathcal{G}$,

\begin{equation}
\mathcal{V}(X)=\frac{1}{6}\ C_{IJK}X^{I}X^{J}X^{K}.  \label{s}
\end{equation}

Non-simple Jordan algebras of degree three are of the form $\mathbb{R\oplus
\Sigma }_{n}$, where $\mathbb{\Sigma }_{n}$ is the Jordan algebra associated
with a quadratic form. The corresponding symmetric scalar manifolds are

\begin{equation}
\mathcal{M}=SO(1,1)\times \frac{SO(n-1,1)}{SO(n-1)}.  \label{scalman}
\end{equation}

In this case, $\mathcal{V}(X)$ is factorizable into a linear times a
quadratic form in $(n-1)$ scalars, which for the positivity of the kinetic
terms in the Lagrangian, must have a Minkowski metric. For simple Jordan
algebras, one obtains four sporadic locally symmetric spaces related to the
four simple unital formally real Jordan algebras over the four division
algebras $\bR$, $\bC$, $\bb{H}$, $\bb{O}$. For more details we
refer the reader to \cite{GST}.

For M-theory compactification on a Calabi-Yau threefold with Hodge numbers 
$h_{(1,1)}$ and $h_{(2,1)},$ the five dimensional theory contains the gravity
multiplet, $h_{(1,1)}-1$ vector multiplets and $h_{(2,1)}+1$
hypermultiplets. The $(h_{(1,1)}-1)$-dimensional space of scalar components
of the abelian vector supermultiplets coupled to supergravity can be
regarded as a hypersurface of a $h_{(1,1)}$-dimensional manifold whose
coordinates $X^{I}(\phi )$ are in correspondence with the vector bosons
(including the graviphoton). The defining equation of the hypersurface is as
in (\ref{s})
\begin{equation}
\mathcal{V}(X)=\frac{1}{6}\ C_{IJK}X^{I}X^{J}X^{K}=X^{I}X_{I}=1,\qquad
I,J,K=1,\ldots,h_{(1,1)}.
\end{equation}
Here $C_{IJK}$ are the topological intersection numbers of the Calabi-Yau, 
$X_{I}$ are the so called ``dual'' special coordinates.

The bosonic part of the ungauged supersymmetric $N=2$ Lagrangian which
describes the coupling of vector multiplets to supergravity is given by
\begin{equation}
e^{-1}\mathcal{L}=\frac{1}{2}R-{\frac{1}{4}}G_{IJ}F_{\mu \nu }{}^{I}F^{\mu
\nu J}-\frac{1}{2}\mathcal{G}_{ij}\partial _{\mu }\phi ^{i}\partial ^{\mu
}\phi ^{j}+{\frac{e^{-1}}{48}}\epsilon ^{\mu \nu \rho \sigma \lambda
}C_{IJK}F_{\mu \nu }^{I}F_{\rho \sigma}^{J}A_{\lambda}^{K}.
\label{generalact}
\end{equation}

The corresponding vector and scalar metric are completely encoded in the
function $\mathcal{V}(X)$,
\begin{align}
G_{IJ}& =-\frac{1}{2}\partial_I\partial_J
\ln \mathcal{V}(X)|_{{\mathcal{V}}=1}\, \notag \\
{\cal G}_{ij}& =G_{IJ}\partial_i X^I\partial_j X^J|_{\mathcal{V}=1}\,
\label{metric}
\end{align}
where $\partial_{i}$ and $\partial_{I}$ refer, respectively, to partial
derivatives with respect to the scalar fields $\phi^{i}$ and
$X^{I}=X^{I}(\phi^{i})$.

Further useful relations are
\begin{equation}
\partial _{i}X_{I}=-{\frac{2}{3}}G_{IJ}\partial _{i}X^{J}\quad ,\quad 
X_{I}={%
\frac{2}{3}}G_{IJ}X^{J}\ .  \label{alsouseful}
\end{equation}
It is worth pointing out that for Calabi-Yau compactifications, $\mathcal{V}$
represents the intersection form, $X^{I}$ and $X_{I}={\frac{1}{6}}%
C_{IJK}X^{J}X^{K}$ correspond, respectively, to the size of the two- and
four-cycles of the Calabi-Yau threefold.

\section{Black Holes in the $STU=1$ Model}
\label{blackholes}

In the last few years considerable progress has been made
in the study of BPS black hole states of the low-energy effective actions of
compactified string and $M$-theory. This was mainly motivated by the
important role that these states play in the understanding of the
non-perturbative structure of string theory. The magnetic and electric BPS
solutions of five-dimensional $N=2$ supergravity models coupled to vector
and hypermultiplets can be regarded as solitons
interpolating between two vacua: Minkowski flat space at infinity and $%
AdS_{3}\times S^{2}$ and $AdS_{2}\times S^{3}$ near the horizon. At a
generic point in space-time, the BPS solution breaks half of supersymmetry.
However, near the horizon supersymmetry is enhanced and fully restored.

In M-theory compactified on a Calabi-Yau threefold, electrically charged
point-like
and magnetically charged string-like BPS states correspond to the two and
five-branes of M-theory wrapped around the two- and four-cycles of the
Calabi-Yau space respectively.
Though the details of the low-energy Lagrangian depend
very much on the geometric and topological data of the compactified
Calabi-Yau space, the analysis of the BPS solutions is considerably
simplified by the rich geometric structure based on ``very special
geometry'' underlying the $N=2$ five-dimensional theories with vector
supermultiplets \cite{sabra1}.

The metric for the BPS black hole solutions can be brought to the form \cite
{chamsabra1}
\begin{eqnarray}
ds^{2} &=&-e^{-4V}(dt+w_{m}dx^{m})^{2}+e^{2V}d\vec{x}^{2},  \notag \\
F_{mn}^{I} &=&\partial_{m}(X^{I}{\cal Q}_{n})-\partial_{n}(X^{I}{\cal Q}_{m}),
\notag \\
F_{tm}^{I} &=&-\partial _{m}(e^{-2V}X^{I}),  \notag \\
e^{2V}X_{I} &=&{\frac{1}{3}}H_{I},  \label{ma}
\end{eqnarray}

where $H_{I}$ are harmonic functions, $H_{I}=h_{I}+{\frac{Q_{I}}{r^{2}}}$, 
$h_{I}$ are constants and $Q_{I}$ denote the electric charges.
Furthermore one has ${\cal Q}_{n}=e^{-2V}w_{n%
\text{ }}$, and the field strength of $w_{m}$ is self-dual. If one defines the
rescaled coordinates $Y^{I}=e^{V}X^{I},$ then the underlying very special
geometry implies that

\begin{equation*}
e^{3V}=\frac{1}{6}\ C_{IJK}Y^{I}Y^{J}Y^{K}.
\end{equation*}

As a general magnetic string solution of $D=5$,
$N=2$ supergravity, one obtains \cite{chamsabra2}

\begin{eqnarray}
ds^{2} &=&e^{-W}(-dt^{2}+dz^{2})+e^{2W}d\vec{x}^{2},  \notag \\
F_{mn}^{I} &=&-\epsilon _{mnp}\partial _{p}H^{I},\text{ \ \ \ }%
e^{W}X^{I}=H^{I},  \notag \\
e^{3W} &=&{\frac{1}{6}}C_{IJK}H^{I}H^{J}H^{K},
\end{eqnarray}

with the harmonic functions
\begin{equation}
H^{I}=h^{I}+{\frac{P^{I}}{r}},
\end{equation}
where $h^{I}$ are constants and $P^{I}$ are magnetic charges.\\
In the
electric case, the near-horizon geometry is given by $AdS_{2}\times S^{3}$ and
the black hole entropy, related to the horizon volume $S^{3}$, is given in
terms of the extremized electric central charge. For the magnetically
charged $D=5$ BPS black string, with near-horizon geometry $AdS_3 \times S^2$,
one similarly finds that the extremized value
of the BPS tension is related to the volume of the $S^{2}$.

As an example we consider the $STU=1$ model \cite{chou,sabra1}. This can be
obtained by compactification of heterotic string theory on $K_{3}\times 
S^{1}
$ \cite{antoniadis}. The tree-level prepotential of this model is given by
\begin{equation}
\mathcal{V}=STU=1,
\end{equation}
and corresponds to the scalar manifold (\ref{scalman}) for $n=2$. Taking $%
S=X^{0}$, $T=X^{1}$ and $U=X^{2}$, one gets for the matrix $G^{IJ}$
\begin{equation}
G^{IJ}=2\left(
\begin{array}{ccc}
S^{2} & 0 & 0 \\
0 & T^{2} & 0 \\
0 & 0 & U^{2}
\end{array}
\right) .
\end{equation}
Considering $S$ as the dependent field, i.~e.~$S=1/(TU)$, we find
\begin{equation}
\mathcal{G}_{ij}=\left(
\begin{array}{cc}
\frac{1}{T^{2}} & \frac{1}{2TU} \\
\frac{1}{2TU} & \frac{1}{U^{2}}
\end{array}
\right) ,\qquad \mathcal{G}^{ij}=\frac{4}{3}\left(
\begin{array}{cc}
T^{2} & -\frac{TU}{2} \\
-\frac{TU}{2} & U^{2}
\end{array}
\right) .
\end{equation}
The field equations following from the action (\ref{generalact}) admit the
non-extremal static black hole solution \cite{bcs}
\begin{eqnarray}
ds^{2} &=&-e^{-4V}fdt^{2}+e^{2V}(f^{-1}dr^{2}+r^{2}d\Omega _{3}^{2}),  
\notag
\\
F_{rt}^{I} &=&-H_{I}^{-2}\partial _{r}\tilde{H}_{I},  \label{staticbh} \\
X^{I} &=&H_{I}^{-1}e^{2V},  \notag
\end{eqnarray}
where
\begin{equation}
d\Omega _{3}^{2}=d\theta ^{2}+\sin ^{2}\theta d\phi ^{2}+\cos ^{2}\theta
d\psi ^{2}  \label{metricS3}
\end{equation}
denotes the metric on the three sphere $S^{3}$. The $H_{I}$ are harmonic
functions given by
\begin{equation}
H_{I}=1+\frac{Q_{I}}{r^{2}},  \label{H_I}
\end{equation}
and $V$ reads
\begin{equation}
e^{2V}=(H_{0}H_{1}H_{2})^{1/3}.  \label{V}
\end{equation}
Furthermore we have
\begin{equation}
f=1-\frac{\mu }{r^{2}}
\end{equation}
with the nonextremality parameter $\mu $, and
\begin{equation}
\tilde{H}_{I}=1+\frac{\tilde{Q}_{I}}{r^{2}},
\end{equation}
where the $\tilde{Q}_{I}$ denote the physical electric charges. They are
related to the $Q_{I}$ appearing in (\ref{H_I}) by the equations
\begin{eqnarray}
Q_{I} &=&\frac{\mu }{2}\sinh \beta _{I}\tanh \frac{\beta _{I}}{2},  \notag 
\\
\tilde{Q}_{I} &=&\frac{\mu }{2}\sinh \beta _{I}.
\end{eqnarray}
The extremal (BPS) limit is reached when $\beta _{I}\rightarrow \infty $, $%
\mu \rightarrow 0$, with $\mu \sinh \beta _{I}$ kept fixed.\newline
For the ADM mass $M_{ADM}$, the Bekenstein-Hawking entropy $S_{BH}$, and the
Hawking temperature $T_{H}$ one obtains
\begin{eqnarray}
M_{ADM} &=&\frac{\pi }{4G_{5}}(\sum_{I}Q_{I}+\frac{3}{2}\mu ),  \label{MADM}
\\
S_{BH} &=&\frac{A_{hor}}{4G_{5}}=\frac{\pi ^{2}}{2G_{5}}\prod_{I}(\mu
+Q_{I})^{1/2},  \label{SBH} \\
T_{H} &=&\frac{\mu }{\pi \prod_{I}(\mu +Q_{I})^{1/2}}.  \label{TH}
\end{eqnarray}
In the extremal case, also a rotating generalization of (\ref{staticbh}) can
be obtained from the general form (\ref{ma}). Its metric is given by
\begin{equation}
ds^{2}=-e^{-4V}(dt+w_{\phi }(r,\theta )d\phi + w_{\psi }(r,\theta )d\psi
)^{2}+e^{2V}(dr^{2}+r^{2}d\Omega ^{2}), \label{rotatingbh}
\end{equation}
where
\begin{eqnarray}
w_{\phi }(r,\theta ) &=&-\frac{\alpha \sin ^{2}\theta }{r^{2}},  \notag \\
w_{\psi }(r,\theta ) &=&\frac{\alpha \cos ^{2}\theta }{r^{2}}.
\end{eqnarray}
The gauge fields are
\begin{equation}
A_{t}^{I}=e^{-2V}X^{I},\qquad A_{\phi }^{I}=e^{-2V}X^{I}w_{\phi },\qquad
A_{\psi }^{I}=e^{-2V}X^{I}w_{\psi }.
\end{equation}
The moduli $X^{I}$ and the functions $V$ and $H_{I}$ are as in (\ref
{staticbh}), (\ref{V}) and (\ref{H_I}) respectively, and the ADM mass is
given by (\ref{MADM}) for $\mu =0$. The Bekenstein-Hawking entropy and the
angular momenta read \cite{chamsabra1}
\begin{eqnarray}
S_{BH} &=&\frac{A_{hor}}{4G}=\frac{\pi ^{2}}{2G_{5}}\sqrt{%
Q_{0}Q_{1}Q_{2}-\alpha ^{2}},  \notag \\
J^{\phi } &=&-J^{\psi }=\frac{\alpha \pi }{4G_{5}}.  \notag
\end{eqnarray}

\section{Near-Horizon Limit and Isometry Superalgebra}

In the following two sections, we shall be
particularly interested in the near-horizon
limit of (\ref{rotatingbh}). For $r\rightarrow 0$ we can write
\begin{equation}
e^{2V}=\frac{Z_{hor}}{3r^{2}},
\end{equation}
where $Z=Q_{I}X^{I}$ is the central charge, and $%
Z_{hor}=3(Q_{0}Q_{1}Q_{2})^{1/3}$ is its value at the horizon. Introducing
the horospherical coordinates $(\tau ,\rho )$,
\begin{equation}
\tau =\frac{t}{2}\sqrt{\frac{3}{Z_{hor}}},\qquad \rho 
=2r^{2}\sqrt{\frac{3}{%
Z_{hor}}},
\end{equation}
one gets for the near-horizon metric
\begin{eqnarray}
ds^{2} &=&-\rho ^{2}d\tau ^{2}+\frac{Z_{hor}}{12\rho ^{2}}d\rho ^{2}+\frac{%
6\alpha }{Z_{hor}}\rho d\tau (\sin ^{2}\theta d\phi -\cos ^{2}\theta d\psi )
\notag \\
&&+\frac{Z_{hor}}{3}\left( d\Omega ^{2}-\frac{27\alpha ^{2}}{Z_{hor}^{3}}%
(\sin ^{2}\theta d\phi -\cos ^{2}\theta d\psi )^{2}\right) .
\label{nearhormetric}
\end{eqnarray}
We observe that, in contrast to the case of vanishing rotation parameter,
the spacetime does not split into a product $AdS_{2}\times S^{3}$. Although
the $AdS_{2}$ part is the same as without rotation, there are nondiagonal
elements, and the three-sphere is distorted. The isometry superalgebra of
the near-horizon supergravity configuration was determined in \cite
{gauntlett1}, where the fact that the residual isometry supergroup can be
determined (modulo bosonic factors) from a knowledge of the Killing spinors
\cite{gauntlett2,townsend,figueroa} has been used. In this way, one obtains
that the near-horizon geometry is invariant under the superalgebra $%
su(1,1|2)\oplus u(1)$ in the rotating case, and under $su(1,1|2)\oplus 
su(2)$
for $\alpha =0$ \cite{gauntlett1}. Thus, for $\alpha \neq 0$, the bosonic
subalgebra is $su(1,1)\oplus su(2)_{L}\oplus u(1)_{R}$. In fact, the
near-horizon spacetime is a homogeneous manifold of the form $[SO(2,1)\times
SU(2)_{L}\times U(1)_{R}]/[U(1)\times U(1)]$ \cite{gauntlett1}. The
conformal algebra $su(1,1)\cong so(2,1)$ is generated by the Killing vectors
\cite{gauntlett1}
\begin{eqnarray}
h &=&\partial _{\tau },  \notag \\
d &=&\rho \partial _{\rho }-\tau \partial _{\tau }, \\
k &=&-\frac{Z_{hor}}{12\rho ^{2}}(1-\frac{27\alpha ^{2}}{Z_{hor}^{3}}%
\partial _{\tau })-\tau ^{2}\partial _{\tau }+2\tau \rho \partial _{\rho }-%
\frac{3\alpha }{2Z_{hor}\rho }(\partial _{\phi }-\partial _{\psi }),  \notag
\end{eqnarray}
satisfying
\begin{equation}
\lbrack d,h]=h,\qquad \lbrack d,k]=-k,\qquad \lbrack h,k]=2d.
\end{equation}
Thus, although the manifold is not a product $AdS_{2}\times S^{3}$, we find
the $so(2,1)$ symmetry inherent to $AdS_{2}$. In the following section, we
will see that this symmetry, which is the conformal symmetry in $0+1$
dimensions, occurs also in the action of a particle charged under the
vectors moving in the near-horizon regime.

\section{Particle Motion near the Horizon}

\label{particle}

We now consider a particle of mass $m$, carrying the charges $q_I$ under the
abelian vectors, which moves in the background (\ref{nearhormetric}). Like
in \cite{claus}, we introduce the new coordinate $q$,
\begin{equation}
\rho = \frac{Z_{hor}}{3q^2}.
\end{equation}
We use a Hamiltonian formalism, and define
\begin{equation}
\mathcal{H} = \frac 12 g^{\mu\nu}(\Pi_{\mu} - q_I A^I_{\mu}) (\Pi_{\nu} -
q_I A^I_{\nu}),
\end{equation}
where the $\Pi_{\mu}$ denote generalized momenta. For our configuration,
this leads to
\begin{eqnarray}
\mathcal{H} &=& -\frac{9q^4}{2Z_{hor}^2}\Pi_{\tau}^2\left(1 - \frac{%
27\alpha^2} {Z_{hor}^3}\right) + \frac{27q^2\alpha}{Z_{hor}^3}\Pi_{\tau} %
\left[\Pi_{\phi} - \Pi_{\psi} + \frac{Z_{hor}^2}{9\alpha} 
q_IX^I_{hor}\right]
\notag \\
& & - \frac 12 (q_IX^I_{hor})^2 + \frac{3q^2}{2Z_{hor}}\Pi_q^2 + 
\frac{3L^2}{%
2Z_{hor}},
\end{eqnarray}
where
\begin{equation}
L^2 = \Pi_{\theta}^2 + \frac{\Pi_{\phi}^2}{\sin^2\theta} + 
\frac{\Pi_{\psi}^2%
}{\cos^2\theta}
\end{equation}
denotes the conserved angular momentum. As the coordinates $\tau$, $\phi$
and $\psi$ are cyclic, the associated conjugate momenta are constants of
motion. If $\mathcal{H}$ solves the mass-shell constraint $2\mathcal{H} =
-m^2$, $-\Pi_{\tau}$ is to be identified with the particle Hamiltonian $H$.
Setting $\Pi_q = p$ and defining $u = pq$, one obtains
\begin{equation}
H = \frac{p^2}{2F(u)} + \frac{mg}{2q^2F(u)},  \label{ham}
\end{equation}
with
\begin{equation}
mg = L^2 + \frac{Z_{hor}}{3}(m^2 - (q_IX^I_{hor})^2),
\end{equation}
and the function $F(u)$ given by
\begin{equation}
F(u) = \frac 12 \left[\sqrt{C^2 + (1 - \frac{27\alpha^2}{Z_{hor}^3}) 
(\frac{3%
}{Z_{hor}}(u^2+L^2) + m^2 - (q_IX^I_{hor})^2)} + C\right].  \label{F}
\end{equation}
In (\ref{F}), the constant $C$ is defined by
\begin{equation}
C = \frac{9\alpha}{Z_{hor}^2}(\Pi_{\phi} - \Pi_{\psi}) + q_IX^I_{hor}.
\end{equation}
One observes that in the limit
\begin{equation}
Z_{hor} \to \infty, \qquad m - q_IX^I_{hor} \to 0,
\end{equation}
with $Z_{hor}(m - q_IX^I_{hor})$ kept fixed, we have $F(u) \to m$, and (\ref
{ham}) reduces to the DFF model \cite{dff}
\begin{equation}
H = \frac{p^2}{2m} + \frac{g}{2q^2}.
\end{equation}
Note that also the general Hamiltonian (\ref{ham}) describes a model of
conformal mechanics. To see this, we write it in the form
\begin{equation}
H = \frac{p^2}{2f(u)},
\end{equation}
with
\begin{equation}
f(u) = \frac{u^2F(u)}{u^2+mg}.
\end{equation}
The generators of the conformal group are then given by \cite{kumar}
\begin{equation}
H = \frac{p^2}{2f}, \qquad D = \frac 12 u, \qquad K = \frac 12 q^2f,
\end{equation}
satisfying the Poisson bracket algebra
\begin{equation}
[D,H] = H, \qquad [D,K] = -K, \qquad [H,K] = -2D.
\end{equation}

\section{Statistical Entropy from $AdS_2$ Central Charge}

\label{statentr}

In order to determine the central charge of the boundary CFT, we proceed
along the lines of \cite{cadoni,ckz,navarro}, and reduce the bosonic part of
the $D=5$, $N=2$ supergravity action to two dimensions. In this section we
shall only consider nonrotating black holes carrying electric charge.
This means that we can consistently truncate the Chern-Simons term in
(\ref{generalact}), so that the bosonic part of the action in five
dimensions reads
\begin{equation}
I = \frac{1}{16\pi G_5}\int d^5x \sqrt{-g}\left[R - \frac12
G_{IJ}F^I_{\mu\nu} F^{J\mu\nu} - \mathcal{G}_{ij}\partial_{\mu}\phi^i%
\partial^{\mu}\phi^j\right],  \label{action5d}
\end{equation}
where $G_5 = l^3$ denotes Newton's constant.
The matrices $G_{IJ}$ and $\mathcal{G}_{ij}$ for the $STU$ model were given
in section \ref{blackholes}. The reduction ansatz for the metric is
\begin{equation}
ds^2 = ds_{(2)}^2 + l^2\Phi^2 d\Omega^2,
\end{equation}
where $\Phi$ denotes the dilaton and $d\Omega^2$ is given by 
(\ref{metricS3}%
).

One now assumes that the gauge fields, scalars and dilaton do not depend on
the coordinates on the internal $S^{3}$. In this way, one arrives at the
two-dimensional effective action
\begin{equation}
I=\frac{\Omega }{16\pi }\int d^{2}x\sqrt{-g}\left[ \Phi ^{3}R+6\Phi (\nabla
\Phi )^{2}+\frac{6\Phi }{l^{2}}-\Phi ^{3}\mathcal{G}_{ij}\partial _{\alpha
}\phi ^{i}\partial ^{\alpha }\phi ^{j}-\frac{1}{2}\Phi ^{3}G_{IJ}F_{\alpha
\beta }^{I}F^{J\alpha \beta }\right] ,  \label{action2d}
\end{equation}
where $\Omega =2\pi ^{2}$ denotes the volume of the unit $S^{3}$, and early
greek indices $\alpha ,\beta ,\ldots $ refer to two-dimensional spacetime.
We now wish to integrate out the field strength $F_{\alpha \beta }^{I}$
(which in two dimensions must be a multiple of the volume form $\epsilon
_{\alpha \beta }$) from the action. This can be done using the Lagrange
multiplier method of \cite{kaloper}. Let us briefly sketch how this works:
Instead of looking at the gauge field action
\begin{equation}
I_{g}=-\frac{\Omega }{16\pi }\int d^{2}x\sqrt{-g}\frac{1}{2}\Phi
^{3}G_{IJ}F_{\alpha \beta }^{I}F^{J\alpha \beta },
\end{equation}
one looks at the formally extended action
\begin{equation}
\tilde{I}_{g}=\frac{\Omega }{16\pi }\int d^{2}x\left[ -\frac{1}{2}\sqrt{-g}%
\Phi ^{3}G_{IJ}F_{\alpha \beta }^{I}F^{J\alpha \beta }+\lambda
_{I}(F_{\alpha \beta }^{I}-\partial _{\alpha }A_{\beta }^{I}+\partial
_{\beta }A_{\alpha }^{I})\epsilon ^{\alpha \beta }\right] ,
\label{actgaugef}
\end{equation}
where the definition of $F^{I}$ as a field strength associated with $A^{I}$
is implemented by means of the Lagrange multiplier $\lambda _{I}$. Note that
the three variables $F^{I},A^{I}$ and $\lambda _{I}$ are considered as
independent in this setting. Variation with respect to $A^{I}$ yields
\begin{equation}
\partial _{\beta }(\lambda _{I}\epsilon ^{\alpha \beta })=0,  \label{varAI}
\end{equation}
so that $\Lambda _{I}:=\lambda _{I}/\sqrt{-g}$ are constants. Due to (\ref
{varAI}), the term $\lambda _{I}(-\partial _{\alpha }A_{\beta }^{I}+\partial
_{\beta }A_{\alpha }^{I})\epsilon ^{\alpha \beta }$ in the action (\ref
{actgaugef}) is a boundary term and can be dropped. We can then integrate
out the field strength $F^{I}$, using its equation of motion
\begin{equation}
-\Phi ^{3}G_{IJ}F^{J\alpha \beta }+\Lambda _{I}\epsilon ^{\alpha \beta }=0.
\end{equation}
This yields
\begin{equation}
\tilde{I}_{g}=-\frac{\Omega }{16\pi }\int d^{2}x\sqrt{-g}\frac{G^{IJ}\Lambda
_{I}\Lambda _{J}}{\Phi ^{3}},
\end{equation}
so that we are left with the total two-dimensional action
\begin{equation}
I=\frac{\Omega }{16\pi }\int d^{2}x\sqrt{-g}\left[ \Phi ^{3}R+6\Phi (\nabla
\Phi )^{2}+\frac{6\Phi }{l^{2}}-\Phi ^{3}\mathcal{G}_{ij}\partial _{\alpha
}\phi ^{i}\partial ^{\alpha }\phi ^{j}-\frac{G^{IJ}\Lambda _{I}\Lambda 
_{J}}{%
\Phi ^{3}}\right] .  \label{action2dnew}
\end{equation}
The dilaton kinetic term in (\ref{action2dnew}) can be eliminated by a
conformal rescaling
\begin{equation}
\bar{g}_{\alpha \beta }=\Phi ^{2}g_{\alpha \beta }.
\end{equation}
Defining $\bar{\Phi}=\Phi ^{3}$, we obtain
\begin{equation}
I=\frac{\Omega }{16\pi }\int d^{2}x\sqrt{-\bar{g}}\left[ \bar{\Phi}\bar{R}+%
\frac{6}{l^{2}\bar{\Phi}^{1/3}}-\bar{\Phi}\mathcal{G}_{ij}\partial _{\alpha
}\phi ^{i}\partial ^{\alpha }\phi ^{j}-\frac{G^{IJ}\Lambda _{I}\Lambda 
_{J}}{%
\bar{\Phi}^{5/3}}\right] .
\end{equation}
Let us now consider the nonextremal black hole solution (\ref{staticbh}) of
the action (\ref{action5d}), and expand it near extremality. To this end, we
introduce an expansion parameter $\epsilon $ ($\epsilon \rightarrow 0$), and
set
\begin{eqnarray}
t &=&\frac{\tilde{t}}{\epsilon },\qquad r=\sqrt{\epsilon }\tilde{r},\qquad
\mu =\mu _{0}\epsilon ,  \notag \\
\bar{\Phi} &=&\bar{\Phi}_{0}+\epsilon \varphi ,\qquad \phi ^{i}=\phi
_{0}^{i}+\epsilon \tilde{\phi}^{i},  \label{expansion}
\end{eqnarray}
where
\begin{equation}
\bar{\Phi}_{0}=\frac{(\tilde{Q}_{0}\tilde{Q}_{1}\tilde{Q}_{2})^{1/2}}{l^{3}}%
,\qquad \phi 
_{0}^{i}=\tilde{Q}_{i}^{-1}(\tilde{Q}_{0}\tilde{Q}_{1}\tilde{Q}%
_{2})^{1/3}.
\end{equation}
Introducing the new coordinate
\begin{equation}
x=\frac{(\tilde{Q}_{0}\tilde{Q}_{1}\tilde{Q}_{2})^{1/6}}{2l^{2}}(\tilde{r}%
^{2}-\frac{\mu _{0}}{2}),
\end{equation}
we arrive at
\begin{equation}
d\bar{s}^{2}=-(\lambda ^{2}x^{2}-a^{2})d\tilde{t}^{2}+(\lambda
^{2}x^{2}-a^{2})^{-1}dx^{2}  \label{metr2d}
\end{equation}
for the rescaled two-dimensional metric, with $\lambda $ and $a$ given by
\begin{eqnarray}
\lambda  &=&\frac{2l}{(\tilde{Q}_{0}\tilde{Q}_{1}\tilde{Q}_{2})^{1/3}}, \\
a^{2} &=&\frac{\mu _{0}^{2}}{4l^{2}(\tilde{Q}_{0}\tilde{Q}_{1}\tilde{Q}%
_{2})^{1/3}}.
\end{eqnarray}
Defining a new dilaton $\eta $ by
\begin{equation}
\eta =\frac{\Omega \epsilon \varphi }{8\pi },
\end{equation}
we obtain for the action at lowest order in the expansion parameter $%
\epsilon $,
\begin{equation}
I=\frac{1}{2}\int d^{2}x\sqrt{-\bar{g}}\eta \lbrack \bar{R}+2\lambda ^{2}],
\label{JT}
\end{equation}
so the leading order is governed by the Jackiw-Teitelboim (JT) model \cite
{JT}. (\ref{metr2d}), together with the linear dilaton
\begin{eqnarray}
\eta  &=&\eta _{0}\lambda x, \\
\eta _{0} &=&\frac{\Omega \epsilon }{16\pi 
l^{2}}(\tilde{Q}_{0}\tilde{Q}_{1}%
\tilde{Q}_{2})^{2/3}\sum_{I}\tilde{Q}_{I}^{-1},  \notag
\end{eqnarray}
represents a black hole solution of this model \cite{cadmig}, with mass and
thermodynamical entropy given by
\begin{eqnarray}
M_{(2)} &=&\frac{1}{2}\eta _{0}a^{2}\lambda ,  \notag \\
S_{(2)} &=&2\pi \eta _{hor}=2\pi \eta _{0}a.  \label{MS2}
\end{eqnarray}
This black hole spacetime has constant curvature, i.~e.~it is locally $%
AdS_{2}$. Now it is known that the asymptotic symmetries of two-dimensional
anti-de~Sitter space form a Virasoro algebra \cite{cadmig}, similar to the
case of $AdS_{3}$, where one has two copies of Virasoro algebras as
asymptotic symmetries \cite{brown}. When realized canonically in the
Hamiltonian formulation of JT gravity, this algebra was shown to exhibit a
central charge \cite{cadmig,cadmig2}
\begin{equation}
c=24\eta _{0}.  \label{centr}
\end{equation}
Using this central charge in Cardy's formula, the authors of \cite{cadmig}
were able to give a microscopic derivation of the entropy of the
two-dimensional black holes (\ref{metr2d}) in the JT model. Our aim is now
to perform a similar calculation for the near-extremal five dimensional
black hole under consideration, making use of the fact that the
dimensionally reduced supergravity action coincides with the JT model at
leading order in the nonextremality parameter, and that the relevant
two-dimensional metric is given by (\ref{metr2d})\footnote{%
Cf.~\cite{cadoni} for similar computations in the case of heterotic 4D
string black holes.}. First of all, we expand the ADM mass $M_{ADM}$ (\ref
{MADM}) and Bekenstein-Hawking entropy $S_{BH}$ (\ref{SBH}) of the black
hole (\ref{staticbh}) in five dimensions for $\mu \rightarrow 0$, yielding
\begin{eqnarray}
M_{ADM} &=&\frac{\pi }{4l^{3}}\sum_{I}\tilde{Q}_{I}(1+\frac{\mu ^{2}}{8%
\tilde{Q}_{I}^{2}}), \\
S_{BH} &=&\frac{\Omega }{4l^{3}}(\tilde{Q}_{0}\tilde{Q}_{1}\tilde{Q}%
_{2})^{1/2}(1+\frac{\mu }{4}\sum_{I}\tilde{Q}_{I}^{-1}),
\end{eqnarray}
so the small excitations above extremality have the energy
\begin{equation}
\Delta M_{ADM}=\frac{\pi \mu ^{2}}{32l^{3}}\sum_{I}\tilde{Q}_{I}^{-1}
\end{equation}
and entropy
\begin{equation}
\Delta S_{BH}=\frac{\Omega \mu 
}{16l^{3}}(\tilde{Q}_{0}\tilde{Q}_{1}\tilde{Q}%
_{2})^{1/2}\sum_{I}\tilde{Q}_{I}^{-1}.
\end{equation}
Comparing this with the two-dimensional results (\ref{MS2}), one finds $%
\Delta S_{BH}=S_{(2)}$ and $\Delta M_{ADM}=\epsilon M_{(2)}$. The factor $%
\epsilon $ appearing in the relation between the two masses stems from the
fact that $M_{ADM}$ was computed with respect to the Killing vector $%
\partial _{t}$, whereas $M_{(2)}$ is related to $\partial _{\tilde{t}%
}=\epsilon \partial _{t}$. This means that up to these normalizations the
five- and two-dimensional energies and entropies match. Expanding also the
Hawking temperature (\ref{TH}) for small values of the nonextremality
parameter $\mu $, one finds for the temperature dependence of $\Delta 
M_{ADM}
$
\begin{equation}
\Delta M_{ADM}=\frac{\pi ^{3}T_{H}^{2}}{32l^{3}}\tilde{Q}_{0}\tilde{Q}_{1}%
\tilde{Q}_{2}\sum_{I}\tilde{Q}_{I}^{-1},
\end{equation}
so the energy of the excitations above extremality is that of an ideal gas
of massless particles in $1+1$ dimensions. This suggests that the
microstates should be described by a two-dimensional field theory rather
than a quantum mechanical system. Let us now proceed with the computation of
the statistical entropy, using the central charge (\ref{centr}). The
Virasoro generator $L_{0}$ for the black hole (\ref{metr2d}) is given by
\cite{cadmig}
\begin{equation}
L_{0}=\frac{M_{(2)}}{\lambda }=\frac{\Omega \epsilon \mu _{0}^{2}}{128\pi
l^{4}}(\tilde{Q}_{0}\tilde{Q}_{1}\tilde{Q}_{2})^{1/3}\sum_{I}\tilde{Q}%
_{I}^{-1}.
\end{equation}
Inserting this together with the central charge (\ref{centr}) into Cardy's
formula, we get for the statistical entropy
\begin{equation}
S_{stat}=2\pi \sqrt{\frac{cL_{0}}{6}}=\frac{\Omega \mu }{8\sqrt{2}l^{3}}(%
\tilde{Q}_{0}\tilde{Q}_{1}\tilde{Q}_{2})^{1/2}\sum_{I}\tilde{Q}_{I}^{-1},
\end{equation}
which agrees, up to a factor $\sqrt{2}$, with the thermodynamical entropy $%
\Delta S_{BH}$ of the small excitations above extremality. The same mismatch
by a factor $\sqrt{2}$ has been found in \cite{cadmig}, where the authors
proposed an explanation of this for the case when the model (\ref{JT}) comes
from dimensional reduction of three-dimensional $AdS$ gravity. Although in
our case $AdS_{2}$ arises as near-horizon geometry of a higher-dimensional
black hole with no intermediate $AdS_{3}$ geometry involved, we shall see in
the next section that by means of a duality transformation the near-horizon
geometry $AdS_{2}\times S^{3}$ of the extremal black hole becomes $%
AdS_{3}\times S^{2}$. We will then be able to use Strominger's counting of
microstates \cite{strominger1} in order to reproduce correctly the
Bekenstein-Hawking entropy of the black hole.

\section{Duality Invariance of the Supergravity Action}

In this section we will show that in presence of a Killing vector field $%
\partial _{z}$, the supergravity action (\ref{action5d}) is invariant under
a certain generalization of T-duality\footnote{%
By considering (\ref{action5d}) we assumed that the Chern-Simons term does
not contribute. One can easily generalize the discussion below to
nonvanishing CS term. This results in a $\theta $ term in four dimensions,
which does not spoil the considered duality invariance.}. The key
observation is then that the three sphere $S^{3}$ appearing in the black
hole geometry can be written as a Hopf fibration, i.~e.~as an $S^{1}$ bundle
over $\bC P^{1}\approx S^{2}$. Performing then a duality transformation
along the Hopf fibre untwists the $S^{3}$, and transforms the electrically
charged black hole into a magnetically charged black string, which has $%
AdS_{3}\times S^{2}$ as near-horizon limit in the extremal case.\newline
To begin with, we reduce the action (\ref{action5d}) to four dimensions,
using the usual Kaluza-Klein reduction ansatz for the five-dimensional
metric,
\begin{equation}
ds^{2}=e^{k/\sqrt{3}}ds_{4}^{2}+e^{-2k/\sqrt{3}}(dz+\mathcal{A}_{\alpha
}dx^{\alpha })^{2},  \label{KK}
\end{equation}
where $k$ denotes the dilaton, and early greek indices $\alpha ,\beta
,\ldots $ refer to four-dimensional spacetime. Assuming that the fields
appearing in (\ref{action5d}) are independent of $z$, one arrives at the
four-dimensional action
\begin{equation}
I_{4}=\frac{L}{16\pi G_{5}}\int d^{4}x\sqrt{-g_{4}}\left[ R_{4}-\frac{1}{2}%
(\nabla k)^{2}-\frac{1}{4}e^{-\sqrt{3}k}\mathcal{F}^{2}-\frac{1}{2}e^{-k/%
\sqrt{3}}F^{2}-\mathcal{G}_{ij}\partial _{\alpha }\phi ^{i}\partial ^{\alpha
}\phi ^{j}\right] ,  \label{action4d}
\end{equation}
where $L$ denotes the length of the circle parametrized by $z$, 
$\mathcal{F}$
is the field strength associated to the Kaluza-Klein vector potential $%
\mathcal{A}$, and
\begin{equation}
\mathcal{F}^{2}=\mathcal{F}_{\alpha \beta }\mathcal{F}^{\alpha \beta
},\qquad F^{2}=G_{IJ}F_{\alpha \beta }^{I}F^{J\alpha \beta }.
\end{equation}
We now dualize both $\mathcal{F}$ and $F^{I}$, using again the Lagrange
multiplier method of \cite{kaloper}. Dropping boundary terms, we arrive at
the dualized action
\begin{eqnarray}
I_{4} &=&\frac{L}{16\pi G_{5}}\int d^{4}x\sqrt{-g_{4}}[R_{4} -
\frac{1}{2}(\nabla k)^{2}-\frac{1}{4}e^{\sqrt{3}k}(^{\star }\mathcal{F})^{2}
\notag \\
&&- \frac{1}{2}e^{k/\sqrt{3}}\frac{1}{4}G^{IJ}\,^{\star }F_{I\alpha
\beta }\,^{\star }F_{J}^{\alpha \beta }-\mathcal{G}_{ij}\partial _{\alpha
}\phi ^{i}\partial ^{\alpha }\phi ^{j}] ,  \label{dualaction4d}
\end{eqnarray}
where we defined
\begin{eqnarray}
^{\star }\mathcal{F}_{\alpha \beta } &=&\frac{1}{2}e^{-\sqrt{3}k}\epsilon
_{\alpha \beta \gamma \delta }\mathcal{F}^{\gamma \delta }, \\
^{\star }F_{I\alpha \beta } &=&e^{-k/\sqrt{3}}G_{IJ}\epsilon _{\alpha \beta
\gamma \delta }F^{J\gamma \delta }.
\end{eqnarray}
Comparing (\ref{dualaction4d}) with (\ref{action4d}), we observe that the
gravitational and gauge field parts of the four-dimensional action, as well
as the dilaton kinetic energy, are invariant under the $\bZ_{4}$
transformation
\begin{equation}
k\rightarrow -k,\qquad \mathcal{F}_{\alpha \beta }\rightarrow \,^{\star }%
\mathcal{F}_{\alpha \beta },\qquad F_{\alpha \beta }^{I}\rightarrow
\,^{\star }F_{I\alpha \beta },\qquad G_{IJ}\rightarrow \frac{1}{4}G^{IJ}.
\label{duality}
\end{equation}
The $\bZ_{4}$ is actually a subgroup of the usual symplectic $Sp(2m+2,\bR)$
duality group \cite{gaillard,craps} of $D=4$, $N=2$ supergravity (coupled to
$m$ vector multiplets) generated by
\begin{equation}
S=\left(
\begin{array}{cc}
{\mathbf{0}} & {\mathbf{1}} \\
{\mathbf{-1}} & {\mathbf{0}}
\end{array}
\right) .
\end{equation}
Note that the transformation $G_{IJ}\rightarrow G^{IJ}/4$ means that
\begin{eqnarray}
X^{I} &\rightarrow &3X_{I}=\frac{1}{2}C_{IJK}X^{J}X^{K},  \notag \\
X_{I} &\rightarrow &\frac{1}{3}X^{I},  \label{dualcoord}
\end{eqnarray}
so essentially the special coordinates go over into their duals. The fact
that this dualization implies $G_{IJ}\rightarrow G^{IJ}/4$ can be shown
using the expression
\begin{equation}
G_{IJ}=\frac{9}{2}X_{I}X_{J}-\frac{1}{2}C_{IJK}X^{K},
\end{equation}
as well as the ''adjoint identity''
\begin{equation}
C_{IJK}C_{J^{\prime }\left( LM\right. }C_{\left. PQ\right) K^{\prime
}}\delta ^{JJ^{\prime }}\delta ^{KK^{\prime }}=\frac{4}{3}\delta _{I\left(
L\right. }C_{\left. MPQ\right) }
\end{equation}
of the associated Jordan algebra \cite{GST}. It can also be seen that this
duality transformation is consistent with the relations (\ref{alsouseful}).
Furthermore, making use of the equation
\begin{equation}
\mathcal{G}_{ij}\partial _{\alpha }\phi ^{i}\partial ^{\alpha }\phi
^{j}=G_{IJ}\partial _{\alpha }X^{I}\partial ^{\alpha }X^{J},
\end{equation}
one checks that (\ref{dualcoord}) does not change the kinetic term of the
scalar fields, so (\ref{duality}), (\ref{dualcoord}) represent in fact a
duality invariance of the four-dimensional action (\ref{action4d}). In the
special case of the $STU=1$ model, (\ref{dualcoord}) implies that the moduli
$\phi ^{i}$ go over into their inverse,
\begin{equation}
\phi ^{i}\rightarrow \frac{1}{\phi ^{i}}.
\end{equation}
We now wish to apply the duality (\ref{duality}), (\ref{dualcoord}) to the
black hole solution (\ref{staticbh}). To this end, we consider the $S^{3}$
as an $S^{1}$ bundle over $S^{2}$, and write for its metric
\begin{equation}
d\Omega ^{2}=\frac{1}{4}\left[ d\vartheta ^{2}+\sin ^{2}\vartheta d\varphi
^{2}+(d\zeta +\cos \vartheta d\varphi )^{2}\right] ,
\end{equation}
where $\zeta $ ($0\leq \zeta \leq 4\pi $) parametrizes the $S^{1}$ fibre.
Introducing the coordinate $z=\lambda \zeta $, where $\lambda $ denotes an
arbitrary length scale, one can write the 5d metric in the KK form 
(\ref{KK}%
), where
\begin{eqnarray}
ds_{4}^{2} &=&\frac{re^{V}}{2\lambda }\left[
-e^{-4V}fdt^{2}+e^{2V}f^{-1}dr^{2}+e^{2V}\frac{r^{2}}{4}(d\vartheta
^{2}+\sin ^{2}\vartheta d\varphi ^{2})\right] ,  \notag \\
e^{-k/\sqrt{3}} &=&\frac{re^{V}}{2\lambda }, \\
\mathcal{A} &=&\lambda \cos \vartheta d\varphi .  \notag
\end{eqnarray}
(Note that $\mathcal{F}=d\mathcal{A}$ is essentially the K\"{a}hler form on 
$%
S^{2}$). We now dualize in 4d according to (\ref{duality}), and then relift
the solution to five dimensions. This yields the configuration
\begin{eqnarray}
ds^{2} &=&e^{-2V}\left[ \frac{\mu }{4\lambda ^{2}}dt^{2}+2dzdt+\frac{%
4\lambda ^{2}}{r^{2}}dz^{2}\right] +\frac{r^{2}}{4\lambda ^{2}}e^{4V}\left[
f^{-1}dr^{2}+\frac{r^{2}}{4}d\Omega _{2}^{2}\right] ,  \notag \\
F_{\vartheta \varphi }^{I} &=&\frac{\tilde{Q}_{I}}{4\lambda }\sin \vartheta 
,
\label{magnblstr} \\
X^{I} &=&H_{I}e^{-2V}.  \notag
\end{eqnarray}
One effect of the duality transformation is thus the untwisting of the Hopf
fibration\footnote{%
The fact that Hopf bundles can be untwisted by T-dualities was observed in
\cite{dlp,halyo}. The idea of untwisting and twisting fibres
to relate strings and black holes, and thus to gain new insights into
black hole microscopics, was also explored in \cite{cvetic2}.}.
Although the metric in (\ref{magnblstr}) contains
nondiagonal elements proportional to $dzdt$, there is no rotation present.
To see this, one observes that the nondiagonal elements come from the vector
potential $\mathcal{A}$ in four dimensions, which gives rise to the field
strength $\mathcal{F}$. The equations of motion for $\mathcal{F}$ following
from the action (\ref{action4d}) read
\begin{equation}
\nabla _{\alpha }(e^{-k\sqrt{3}}\mathcal{F}^{\alpha \beta })=0,
\end{equation}
so there exists an associated conserved charge
\begin{equation}
J=\int_{S_{\infty }^{2}}d^{2}S_{\alpha \beta }e^{-k\sqrt{3}}\mathcal{F}%
^{\alpha \beta }.
\end{equation}
For the solution (\ref{magnblstr}) under consideration, however, one easily
verifies that $J$ (which, up to a normalization factor, represents the
angular momentum) vanishes.\newline
One can further simplify (\ref{magnblstr}) by an $SL(2,\bR)$ transformation
\begin{equation}
\left(
\begin{array}{c}
t^{\prime } \\
z^{\prime }
\end{array}
\right) =\left(
\begin{array}{cc}
0 & -\frac{2\lambda }{\sqrt{\mu }} \\
\frac{\sqrt{\mu }}{2\lambda } & \frac{2\lambda }{\sqrt{\mu }}
\end{array}
\right) \left(
\begin{array}{c}
t \\
z
\end{array}
\right) .  \label{SL2R}
\end{equation}
Introducing also the new radial coordinate $\rho =r^{2}/(4\lambda )$, we
then get for the metric
\begin{equation}
ds^{2}=e^{-2V}(-fdt^{\prime }{}^{2}+dz^{\prime }{}^{2})+e^{4V}(f^{-1}d\rho
^{2}+\rho ^{2}d\Omega _{2}^{2}).  \label{nonextrblstr}
\end{equation}
(\ref{nonextrblstr}), together with the gauge and scalar fields given in (%
\ref{magnblstr}), represents a nonextremal generalization of the
supersymmetric magnetic black string found in \cite{chamsabra2}. The duality
(\ref{duality}) thus maps electrically charged black holes onto
magnetically charged black strings.\newline
Now a short comment on the $SL(2,\bR)$ transformation (\ref{SL2R}) is in
order. The orbits of the Killing vector
\begin{equation}
\partial _{z^{\prime }}=\frac{2\lambda }{\sqrt{\mu }}\partial _{t}
\end{equation}
are non-compact since the time coordinate is non-compact. This means that
globally the spacetimes in (\ref{magnblstr}) and (\ref{nonextrblstr}) are
not equivalent. To make the transformation (\ref{SL2R}) a symmetry, we have
to compactify the orbits of $\partial _{z^{\prime }}$. We shall see below
however, that the temperature and entropy of one black string can be deduced
from the other, which indicates that the two solutions (\ref{magnblstr}) and
(\ref{nonextrblstr}) are in the same universality class \cite{skenderis}.%
\newline
The Bekenstein-Hawking entropy of the black string (\ref{nonextrblstr})
results to coincide precisely with that of the dual black hole given by 
(\ref
{SBH}), if we assign to $z^{\prime }$ the period $\Delta z2\lambda /\sqrt{%
\mu }$, where $\Delta z=4\pi \lambda $ denotes the period of $z$. The
Hawking temperature can be computed by requiring the absence of conical
singularities in the Euclidean metric, yielding
\begin{equation}
T_{H}=\frac{2\lambda \sqrt{\mu }}{\pi \prod_{I}(\mu +Q_{I})^{1/2}},
\end{equation}
i.~e.~$2\lambda /\sqrt{\mu }$ times the black hole temperature (\ref{TH}).
The factor $2\lambda /\sqrt{\mu }$ stems from the rescaling of the time
coordinate contained in (\ref{SL2R}). Thus, up to this normalization, the
temperature and entropy of the black string (\ref{nonextrblstr}) coincide
with that of the dual black hole (\ref{staticbh}), i.~e.~they are duality
invariant.

\section{Microstate Counting from $AdS_3$ Gravity}

We now want to use the near-horizon geometry of the dual solution (\ref
{nonextrblstr}) to count the microstates giving rise to the
Bekenstein-Hawking entropy. In \cite{chamsabra2} it was shown that in the
extremal case, the geometry becomes $AdS_{3}\times S^{2}$ near the event
horizon. The idea is now to use the central charge of $AdS_{3}$ gravity 
\cite
{brown} in Cardy's formula, in order to compute the statistical entropy,
like it was done by Strominger \cite{strominger1} for the BTZ black
hole\footnote{Cf.~also \cite{cvetic3}, where similar computations for black
strings in six dimensions with $BTZ \times S^3$ near-horizon geometry
were performed.}.
As only the $AdS_{3}$ part is relevant, we would like to reduce the
supergravity action from five to three dimensions. To this end, we first
Hodge-dualize the magnetic two-form field strength in (\ref{magnblstr}).
This yields for the action (\ref{action5d})
\begin{equation}
I=\frac{1}{16\pi G_{5}}\int d^{5}x\sqrt{-g}\left[ R-\frac{1}{2}G^{IJ}H_{I\mu
\nu \rho }H_{J}^{\mu \nu \rho }-\mathcal{G}_{ij}\partial _{\mu }\phi
^{i}\partial ^{\mu }\phi ^{j}\right] ,  \label{action5dhodge}
\end{equation}
where
\begin{equation}
H_{I\mu \nu \rho }=-\frac{1}{2\sqrt{3}}G_{IJ}\epsilon _{\mu \nu \rho \lambda
\sigma }F^{J\lambda \sigma }.
\end{equation}
Note that for the solution under consideration, the $H_{I}$ do not depend on
the coordinates of the internal $S^{2}$. Furthermore, in 3d the three-forms 
$%
H_{I}$ are proportional to the volume form and can be integrated out. For
the metric, we use the reduction ansatz
\begin{equation}
ds^{2}=ds_{3}^{2}+l^{2}\Phi ^{2}d\Omega _{2}^{2},
\end{equation}
where $G_{5}=l^{3}$ as before, and $d\Omega _{2}^{2}$ denotes the standard
metric on the unit $S^{2}$. This gives the reduced action
\begin{equation}
I=\frac{1}{4l}\int d^{3}x\sqrt{-g_{3}}\Phi ^{2}\left[ R_{3}+\frac{2}{%
l^{2}\Phi ^{2}}+\frac{2}{\Phi ^{2}}(\nabla \Phi )^{2}-\frac{1}{2}%
G^{IJ}H_{I\alpha \beta \gamma }H_{J}^{\alpha \beta \gamma }-\mathcal{G}%
_{ij}\partial _{\alpha }\phi ^{i}\partial ^{\alpha }\phi ^{j}\right] ,
\end{equation}
where early greek indices $\alpha ,\beta ,\ldots $ refer to
three-dimensional spacetime. Using the procedure described in section \ref
{statentr}, the three-forms $H_{I}$ can be integrated out. In this way, one
finally obtains
\begin{equation}
I=\frac{1}{4l}\int d^{3}x\sqrt{-g_{3}}\Phi ^{2}\left[ R_{3}+\frac{2}{%
l^{2}\Phi ^{2}}+\frac{2}{\Phi ^{2}}(\nabla \Phi 
)^{2}-\frac{G_{IJ}P^{I}P^{J}%
}{\Phi ^{4}l^{4}}-\mathcal{G}_{ij}\partial _{\alpha }\phi ^{i}\partial
^{\alpha }\phi ^{j}\right] ,
\end{equation}
where we introduced the magnetic charges
\begin{equation}
P^{I}=\frac{\tilde{Q}_{I}}{4\lambda }
\end{equation}
of the black string (\ref{magnblstr}).\newline
We find it convenient to conformally rescale the metric,
\begin{equation}
\bar{g}_{\alpha \beta }=\Phi g_{\alpha \beta },
\end{equation}
yielding
\begin{equation}
I=\frac{1}{4l}\int d^{3}x\sqrt{-\bar{g}_{3}}\Phi ^{\frac{3}{2}}\left[ 
\bar{R}%
_{3}+\frac{2}{l^{2}\Phi ^{3}}-\frac{3}{2\Phi ^{2}}(\nabla \Phi )^{2}-\frac{%
G_{IJ}P^{I}P^{J}}{\Phi ^{5}l^{4}}-\mathcal{G}_{ij}\partial _{\alpha }\phi
^{i}\partial ^{\alpha }\phi ^{j}\right]   \label{action3d}
\end{equation}
for the action. The conformally rescaled 3d metric reads
\begin{equation}
d\bar{s}_{3}^{2}=\frac{\rho }{l}(-fdt^{\prime }{}^{2}+dz^{\prime
}{}^{2})+e^{6V}\frac{\rho d\rho ^{2}}{lf}.
\end{equation}
The idea is now to expand this metric near the horizon and near extremality.
This can be done by setting
\begin{equation}
t^{\prime }=\frac{t^{\prime \prime }}{\sqrt{\epsilon }}(2\lambda 
)^{4}\sqrt{%
\frac{l}{\mu _{0}\lambda \tilde{Q}_{0}\tilde{Q}_{1}\tilde{Q}_{2}}},\qquad
z^{\prime }=\frac{z^{\prime \prime }}{\sqrt{\epsilon }}\frac{(2\lambda 
)^{2}%
}{\sqrt{\mu _{0}}},\qquad \rho =\epsilon \tilde{r}^{2}\frac{\mu _{0}l}{%
(2\lambda )^{4}},\qquad \mu =\mu _{0}\epsilon ,
\end{equation}
and taking the limit $\epsilon \rightarrow 0$. This leads to the metric
\begin{equation}
d\bar{s}_{3}^{2}=-\frac{\tilde{r}^{2}-\tilde{r}_{+}^{2}}{l_{eff}^{2}}%
dt^{\prime \prime }{}^{2}+\tilde{r}^{2}dz^{\prime \prime }{}^{2}+\frac{%
l_{eff}^{2}d\tilde{r}^{2}}{\tilde{r}^{2}-\tilde{r}_{+}^{2}},  \label{BTZ}
\end{equation}
where we introduced
\begin{eqnarray}
\tilde{r}_{+}^{2} &=&\frac{4\lambda ^{3}}{l},  \notag \\
l_{eff}^{2} &=&\frac{\tilde{Q}_{0}\tilde{Q}_{1}\tilde{Q}_{2}}{16l\lambda 
^{3}%
}.
\end{eqnarray}
We recognize (\ref{BTZ}) as the BTZ black hole \cite{banados}, with event
horizon at $\tilde{r}=\tilde{r}_{+}$. One easily verifies that the period of
the coordinate $z^{\prime \prime }$ is $2\pi $. $\Lambda
_{eff}=-1/l_{eff}^{2}$ is the effective cosmological constant. The effective
3d Newton constant can be read off from the action (\ref{action3d}),
yielding
\begin{equation}
\frac{1}{16\pi G_{eff}}=\frac{1}{4l}\Phi _{hor}^{3/2},
\end{equation}
where the subscript indicates that the dilaton $\Phi $ is to be evaluated at
the horizon. In this way, we get
\begin{equation}
\frac{1}{G_{eff}}=\frac{4\pi }{l^{5/2}}e^{3V_{hor}}\rho _{hor}^{3/2}.
\end{equation}
The Bekenstein-Hawking entropy of the BTZ black hole (\ref{BTZ}) is given by
\begin{equation}
S_{(3)}=\frac{A_{hor}}{4G_{eff}}=\frac{\pi ^{2}}{2l^{3}}\prod_{I}(\mu
+Q_{I})^{1/2},
\end{equation}
which, as it should be, equals the entropy (\ref{SBH}) of the
five-dimensional black hole we started with. The BTZ black hole mass 
$M_{(3)}
$ can be computed using the formula
\begin{equation}
\tilde{r}_{+}^{2}=8G_{eff}M_{(3)}l_{eff}^{2},
\end{equation}
which yields
\begin{equation}
M_{(3)}=\frac{\lambda ^{3}}{2lG_{eff}l_{eff}^{2}}.
\end{equation}
We can now apply Strominger's counting of microstates \cite{strominger1} to
reproduce the Bekenstein-Hawking entropy. To this end, one first observes
that the central charge appearing in the asymptotic symmetry algebra of $%
AdS_{3}$ \cite{brown} in our case reads
\begin{equation}
c=\frac{3l_{eff}}{2G_{eff}}.  \label{central3d}
\end{equation}
Furthermore, we have the relations
\begin{eqnarray}
M_{(3)} &=&\frac{1}{l_{eff}}(L_{0}+\bar{L}_{0}), \\
J &=&L_{0}-\bar{L}_{0}
\end{eqnarray}
for the mass and angular momentum. For (\ref{BTZ}) one has $J=0$, so 
$L_{0}=%
\bar{L}_{0}=\frac{1}{2}l_{eff}M_{(3)}$. Plugging this, together with the
central charge (\ref{central3d}), into Cardy's formula
\begin{equation}
S_{stat}=2\pi \sqrt{\frac{cL_{0}}{6}}+2\pi \sqrt{\frac{c\bar{L}_{0}}{6}}
\end{equation}
yields the statistical entropy
\begin{equation}
S_{stat}=\frac{\pi \lambda 
^{3/2}}{l^{1/2}G_{eff}}=%
\frac{\pi ^{2}}{2l^{3}}\prod_{I}(\mu +Q_{I})^{1/2},
\end{equation}
which coincides precisely with the thermodynamical entropy (\ref{SBH}) of
the 5d black hole (\ref{staticbh}).

\section{Final Remarks}

The conclusions we have drawn are valid for general black holes of $D=5$,
$N=2$ supergravities. In particular they apply also to the case of theories
obtained from compactifications on Calabi-Yau spaces. In different contexts
there has been a discussion of dualities \cite{sfetsos,argurio,julia} which
connect various black hole solutions. We have exhibited an explicit duality
transformation which is an invariance of the action: it turns the $%
AdS_2\times S^3$ near horizon geometry into $AdS_3\times S^2$.

Our calculation shows that the correct statistical entropy is given by the
counting of microstates from $AdS_3$, where both $L_0$ and $\bar{L}_0$ are
different from zero. Using instead the central charge of the $AdS_2$
Virasoro algebra, with only right-movers, gives a factor $\sqrt 2$ mismatch
between statistical and thermodynamical entropy. Within the $AdS_2$ approach
we were able (up to the mentioned factor $\sqrt 2$) to capture only the
entropy of the small excitations above extremality, not that of the ground
state itself. The reason for this was the fact that in two dimensions the
Einstein-Hilbert term is a topological invariant, and does not contribute to
the central charge computed in \cite{cadmig}. In the extremal limit $%
\epsilon \to 0$ the $AdS_2$ central charge (\ref{centr}) vanishes, whereas
the central charge (\ref{central3d}) for $AdS_3$ is given by
\begin{equation}
c = \frac{3\pi}{16l^3\lambda^3}\tilde{Q}_0\tilde{Q}_1\tilde{Q}_2.
\end{equation}
This is in agreement with Strominger's observation \cite{strominger2} that
the $AdS_2$ Virasoro algebra is related to the right-moving $AdS_3$ Virasoro
algebra by a topological twist which shifts the central charge to zero.

It might be that the degeneracy of the ground state itself is effectively
captured by a model of conformal quantum mechanics \cite{gibbtown}. However,
our results support the point of view that the excitations above extremality
are described by a two-dimensional conformal field theory \cite
{horowitz,maldstrom}.


\newpage

\begin{thebibliography}{99}
\bibitem{chou}  A.~Chou, R.~Kallosh, J.~Rahmfeld, S.~-J.~Rey, M.~Shmakova,
and W.~K.~Wong, \emph{Critical Points and Phase Transitions in 5D
Compactifications of M-Theory}, Nucl.~Phys.~\textbf{B508} (1997) 147.

\bibitem{gaida}  I.~Gaida, S.~Mahapatra, T.~Mohaupt, and W.~A.~Sabra, 
\emph{%
Black Holes and Flop Transitions in M-Theory on Calabi-Yau Threefolds},
Class.~Quant.~Grav.~\textbf{16} (1999) 419.

\bibitem{antoine}  B.~de Wit and A.~Van Proeyen, \emph{Broken Sigma-Model
Isometries in Very Special Geometry}, Phys.~Lett.~\textbf{B293} (1992) 94.

\bibitem{chamsabra2}  A.~H.~Chamseddine and W.~A.~Sabra, \emph{Calabi-Yau
Black Holes and Enhancement of Supersymmetry in Five Dimensions},
Phys.~Lett.~\textbf{B460} (1999) 63.

\bibitem{cvetic1} M.~Cveti\v{c} and D.~Youm,
\emph{General Rotating Five Dimensional Black Holes of Toroidally
Compactified Heterotic String}, Nucl.~Phys.~{\bf B476} (1996) 118;\\
\emph{Entropy of Non-Extreme Charged Rotating Black Holes in String Theory},
Phys.~Rev.~{\bf D54} (1996) 2612.

\bibitem{bmpv}  J.~C.~Breckenridge, R.~C.~Myers, A.~W.~Peet, and C.~Vafa,
\emph{D-Branes and Spinning Black Holes}, Phys.~Lett.~\textbf{B391} (1997)
93.

\bibitem{krw}  R.~Kallosh, A.~Rajaraman, and W.~K.~Wong, \emph{%
Supersymmetric Rotating Black Holes and Attractors}, Phys.~Rev.~\textbf{D55}
(1997) 3246.

\bibitem{chamsabra1}  A.~H.~Chamseddine and W.~A.~Sabra, \emph{Metrics
Admitting Killing Spinors in Five Dimensions}, Phys.~Lett.~\textbf{B426}
(1998) 36.

\bibitem{cadmig}  M.~Cadoni and S.~Mignemi, \emph{Entropy of $2D$ Black
Holes from Counting Microstates}, Phys.~Rev.~\textbf{D59} (1999) 081501;%
\newline
\emph{Asymptotic Symmetries of $AdS_2$ and Conformal Group in $D=1$},
Nucl.~Phys.~\textbf{B557} (1999) 165.

\bibitem{fragm}  J.~M.~Maldacena, J.~Michelson, and A.~Strominger, \emph{%
Anti-de~Sitter Fragmentation}, JHEP \textbf{9902} (1999) 011.

\bibitem{maldstrom}  J.~M.~Maldacena and A.~Strominger, \emph{Universal
Low-Energy Dynamics for Rotating Black Holes}, Phys.~Rev.~\textbf{D56}
(1997) 4975.

\bibitem{maldacena}  J.~M.~Maldacena, \emph{The Large $N$ Limit of
Superconformal Field Theories and Supergravity}, Adv.~Theor.~Math.~Phys.~%
\textbf{2} (1998) 231.

\bibitem{vafa}  C.~Vafa, \emph{Black Holes and Calabi-Yau Threefolds},
Adv.~Theor.~Math.~Phys.~\textbf{2} (1998) 207.

\bibitem{stromvafa}  A.~Strominger and C.~Vafa, \emph{Microscopic Origin of
the Bekenstein-Hawking Entropy}, Phys.~Lett.~\textbf{B379} (1996) 99.

\bibitem{claus}  P.~Claus, M.~Derix, R.~Kallosh, J.~Kumar, P.~K.~Townsend,
and A.~Van Proeyen, \emph{Black Holes and Superconformal Mechanics},
Phys.~Rev.~Lett.~\textbf{81} (1998) 4553.

\bibitem{gibbtown}  G.~W.~Gibbons and P.~K.~Townsend, \emph{Black Holes and
Calogero Models}, Phys.~Lett.~\textbf{B454} (1999) 187.

\bibitem{GST}  M.~G\"{u}naydin, G.~Sierra, and P.~K.~Townsend, \emph{The
Geometry of $N=2$ Maxwell-Einstein Supergravity and Jordan Algebras},
Nucl.~Phys.~\textbf{B242} (1984) 244.

\bibitem{sabra1}  W.~A.~Sabra, \emph{General BPS Black Holes in Five
Dimensions}, Mod.~Phys.~Lett.~\textbf{A13} (1998) 239.

\bibitem{antoniadis}  I.~Antoniadis, S.~Ferrara, and T.~R.~Taylor, \emph{$%
N=2 $ Heterotic Superstring and its Dual Theory in Five Dimensions},
Nucl.~Phys.~\textbf{B460} (1996) 489.

\bibitem{bcs}  K.~Behrndt, M.~Cveti\v{c}, and W.~A.~Sabra, \emph{Non-Extreme
Black Holes of Five Dimensional $N=2$ AdS Supergravity}, 
Nucl.~Phys.~\textbf{%
B553} (1999) 317.

\bibitem{gauntlett1}  J.~P.~Gauntlett, R.~C.~Myers, and P.~K.~Townsend,
\emph{Black Holes of D=5 Supergravity}, Class.~Quant.~Grav.~\textbf{16}
(1999) 1.

\bibitem{gauntlett2}  J.~P.~Gauntlett, R.~C.~Myers, and P.~K.~Townsend,
\emph{Supersymmetry of Rotating Branes}, Phys.~Rev.~\textbf{D59} (1999)
025001.

\bibitem{townsend}  P.~K.~Townsend, \emph{Killing Spinors, Supersymmetries
and Rotating Intersecting Branes}, hep-th/9901102, to appear in \textit{%
Proceedings of the 22nd Johns Hopkins Workshop, Gothenburg, 1998}.

\bibitem{figueroa}  J.~M.~Figueroa-O'Farrill, \emph{On the Supersymmetries
of Anti-de~Sitter Vacua}, Class.~Quant.~Grav.~\textbf{16} (1999) 2043.

\bibitem{dff}  V.~De Alfaro, S.~Fubini, and G.~Furlan, \emph{Conformal
Invariance in Quantum Mechanics}, Nuovo Cimento \textbf{34A} (1976) 569.

\bibitem{kumar}  J.~Kumar, \emph{Conformal Mechanics and the Virasoro 
Algebra%
}, JHEP \textbf{9904} (1999) 006.

\bibitem{cadoni}  M.~Cadoni, \emph{Dimensional Reduction of 4D Heterotic
String Black Holes}, Phys.~Rev.~\textbf{D60} (1999) 084016.

\bibitem{ckz}  S.~Cacciatori, D.~Klemm, and D.~Zanon, \emph{$w_{\infty}$
Algebras, Conformal Mechanics, and Black Holes}, hep-th/9910065.

\bibitem{navarro}  J.~Navarro-Salas and P.~Navarro, \emph{$AdS_2$/$CFT_1$
Correspondence and Near-Extremal Black Hole Entropy}, hep-th/9910076.

\bibitem{kaloper}  N.~Kaloper, \emph{Topological Mass Generation in Three
Dimensional String Theory}, Phys.~Lett.~\textbf{B320} (1994) 16.

\bibitem{JT}  C.~Teitelboim, in \emph{Quantum Theory of Gravity}, edited by
S.~M.~Christensen (Hilger, Bristol); R.~Jackiw, ibid.

\bibitem{brown}  J.~D.~Brown and M.~Henneaux, \emph{Central Charges in the
Canonical Realization of Asymptotic Symmetries: An Example from
Three-Dimensional Gravity}, Commun.~Math.~Phys.~\textbf{104} (1986) 207.

\bibitem{cadmig2}  M.~Cadoni and S.~Mignemi, \emph{Symmetry Breaking,
Central Charges and the $AdS_2/CFT_1$ Correspondence}, hep-th/0002256.

\bibitem{strominger1}  A.~Strominger, \emph{Black Hole Entropy from
Near-Horizon Microstates}, JHEP \textbf{9802} (1998) 009.

\bibitem{gaillard}  M.~K.~Gaillard and B.~Zumino, \emph{Duality Rotations
for Interacting Fields}, Nucl.~Phys.~\textbf{B193} (1981) 221.

\bibitem{craps}  B.~Craps, F.~Roose, W.~Troost, and A.~Van Proeyen, \emph{%
What is Special K\"ahler Geometry?}, Nucl.~Phys.~\textbf{B503} (1997) 565.

\bibitem{dlp}  M.~J.~Duff, H.~L\"u, and C.~N.~Pope, \emph{$AdS_5 \times S^5$
Untwisted}, Nucl.~Phys.~\textbf{B532} (1998) 181;\newline
\emph{$AdS_3 \times S^3$ (Un)twisted and Squashed, and an $O(2,2;\bZ)$
Multiplet of Dyonic Strings}, Nucl.~Phys.~\textbf{B544} (1999) 145.

\bibitem{halyo}  E.~Halyo, \emph{Supergravity on $AdS_{5/4} \times$ Hopf
Fibrations and Conformal Field Theories}, hep-th/9803193.

\bibitem{cvetic2} M.~Cveti\v{c}, H.~L\"u, and C.~N.~Pope,
\emph{Decoupling Limit, Lens Spaces and Taub-NUT: $D=4$ Black Hole
Microscopics from $D=5$ Black Holes}, Nucl.~Phys.~{\bf B549} (1999) 194.

\bibitem{skenderis}  K.~Skenderis, \emph{Black Holes and Branes in String
Theory}, Erice lecture notes, hep-th/9901050.

\bibitem{cvetic3} M.~Cveti\v{c} and F.~Larsen,
\emph{Near Horizon Geometry of Rotating Black Holes in Five Dimensions},
Nucl.~Phys.~{\bf B531} (1998) 239.

\bibitem{banados}  M.~Ba\~{n}ados, C.~Teitelboim, and J.~Zanelli, \emph{%
Black Hole in Three-Dimensional Spacetime}, Phys.~Rev.~Lett.~\textbf{69}
(1992) 1849.

\bibitem{sfetsos}  K.~Sfetsos and K.~Skenderis, \emph{Microscopic Derivation
of the Bekenstein-Hawking Entropy Formula for Non-Extremal Black Holes},
Nucl.~Phys.~\textbf{B517} (1998) 179.

\bibitem{argurio}  R.~Argurio, F.~Englert, and L.~Houart, \emph{Statistical
Entropy of the Four-Dimensional Schwarzschild Black Hole}, Phys.~Lett.~%
\textbf{B426} (1998) 275.

\bibitem{julia}  B.~Julia, \emph{Superdualities: Below and Beyond 
U-duality}%
, hep-th/0002035.

\bibitem{strominger2}  A.~Strominger, \emph{$AdS_2$ Quantum Gravity and
String Theory}, JHEP \textbf{9901} (1999) 007.

\bibitem{horowitz}  G.~T.~Horowitz and A.~Strominger, \emph{Counting States
of Near-Extremal Black Holes}, Phys.~Rev.~Lett.~\textbf{77} (1996) 2368;%
\newline
G.~T.~Horowitz, D.~A.~Lowe, and J.~M.~Maldacena, \emph{Statistical Entropy
of Nonextremal Four-Dimensional Black Holes and U-Duality}, 
Phys.~Rev.~Lett.~%
\textbf{77} (1996) 430.
\end{thebibliography}
\end{document}